\documentclass[11pt, oneside]{article}   	
\usepackage{geometry}                		
\usepackage[pdftex]{graphicx}
\usepackage{subcaption}
								
\usepackage{amssymb}
\usepackage{amsmath}
\usepackage{listings}
\usepackage{xcolor}
\newcommand{\bi}{\begin{itemize}}
\newcommand{\ei}{\end{itemize}}
\newcommand{\be}{\begin{equation}}
\newcommand{\ee}{\end{equation}}

\begin{document}
\title{Natural Disaster In Canada (2024)}
\author{H. Hao\footnote{This paper represents my personal view only and doesn't represent views of my (former) employers. }}
\date{September 30, 2025}\maketitle
\begin{abstract}
This paper is a follow-up to our earlier study, Natural Disasters in Canada (2017). We analyze the Canadian Disaster Database (CDD) to examine the frequency and severity of various natural disasters over the past 120 years and to identify emerging trends. We generate annual loss distributions for individual disaster types, as well as an aggregate annual loss distribution across all event types. Our analysis provides evidence that Canada is experiencing warmer and wetter conditions and indicates a substantial likelihood of extreme national-level losses.
\end{abstract}
\titlepage
\section{Summary}
This paper is a follow-up to our earlier study, \textit{Natural Disasters in Canada} (2017). Over the past seven years, Canada — the second-largest country in the world — has experienced its share of natural disasters. In this paper, we compile data on natural disasters from the Canadian Disaster Database (CDD), maintained by Public Safety Canada.

As in the earlier study, our objectives are threefold: 
1) to examine whether natural disasters, by type, have become more frequent in Canada over the past seven years; 
2) to assess whether such disasters have become more severe; and 
3) to evaluate whether, overall, the country faces increasing financial losses each year as a result.

Our key findings are as follows:
\begin{itemize}
    \item \textbf{Frequency:} There is no strong evidence that any particular type of natural disaster has become more frequent nationwide in the past seven years (Sections~\ref{trend} and~\ref{mf}).
    \item \textbf{Severity:} There is evidence that Thunderstorm\footnote{Thunderstorm is referred as Storms and Severe Thunderstorms in CDD. Thunderstorm and Wildfire are two of the thirteen disaster event types defined in CDD.} and Wildfire have caused greater financial losses in recent years, while other event types have remained relatively stable (Sections~\ref{trend},~\ref{ms}, and~\ref{is}). It seems Canada is getting warmer and wetter. 
    \item \textbf{Expected annual losses:} At the national level, we observe an increased likelihood of extreme annual financial losses (Section~\ref{aal}).
\end{itemize}
For background information, readers may refer to:
\begin{itemize}
    \item \textbf{Canadian Disaster Database (CDD):} \ref{cdd} introduces the CDD, its data collection criteria, and definitions. The database categorizes 13 types of natural disasters.
    \item \textbf{Probability and simulation:} \ref{mb} outlines the probabilistic framework and simulation procedures used in this study.
    \item \textbf{2024 collection vs 2017 collection}: 2024 collection contains natural disasters occurred between 1900 and 2020, with 899 events and losses of \$34 B. 2017 collection contains disasters between 1900 and 2016 with 789 events and losses of \$22 B.
\end{itemize}

\section{Trend}\label{trend}
In the 2024 collection, the six most frequent disaster categories are Flood (number of events: 336, weight: 37\%), Thunderstorm (141, 16\%), Wildfire (113, 13\%), Winter Storm (89, 10\%), Tornado (51, 6\%), and Drought (46, 5\%). Together, these six categories account for 86\% of all recorded natural disasters.

\begin{figure}[h!]
\centering
\caption{Number of Disasters since 1900}
\label{nol}
\includegraphics[width=0.5\textwidth]{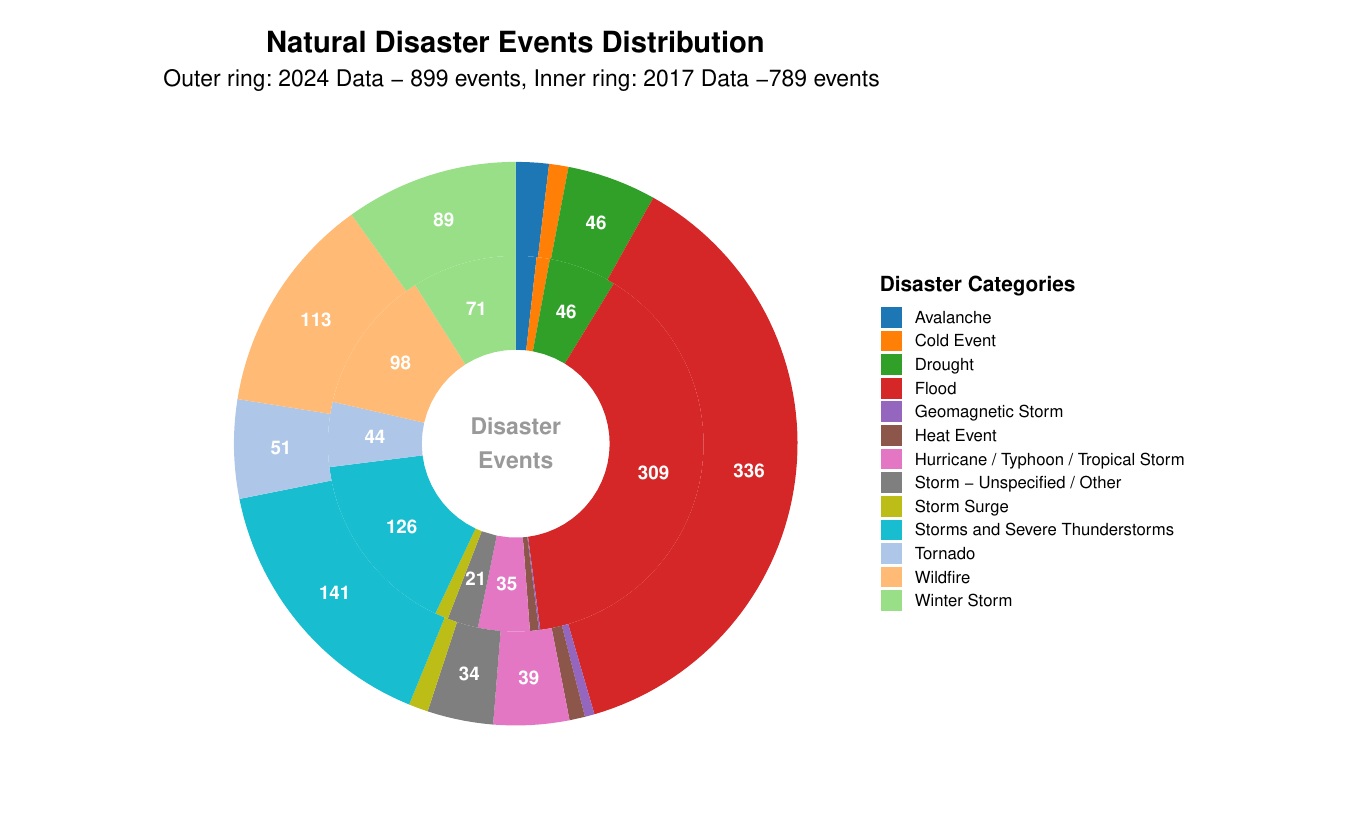}
\end{figure}

In Figure~\ref{nol}, we present the total number of natural disasters from both the 2017 and 2024 collections. Overall, there is no strong evidence that any specific type of natural disaster has become more frequent nationwide over the past seven years.

In terms of total financial losses, the 2024 collection identifies the following top six categories: Flood (total losses: \$10{.}0~B, weight: 31\%), Winter Storm (\$6{.}7~B, 20\%), Thunderstorm (\$5{.}9~B, 17\%), Wildfire (\$4{.}9~B, 15\%), Storms – Unspecified/Other\footnote{Storms – Unspecified/Other is referred as Storm Other hereafter.} (\$2{.}0~B, 6\%), and Drought (\$1{.}7~B, 5\%). These six categories account for 94\% of total disaster-related losses.

\begin{figure}[h!]
\centering
\caption{Losses from Disasters since 1900}
\label{dol}
\includegraphics[width=0.5\textwidth]{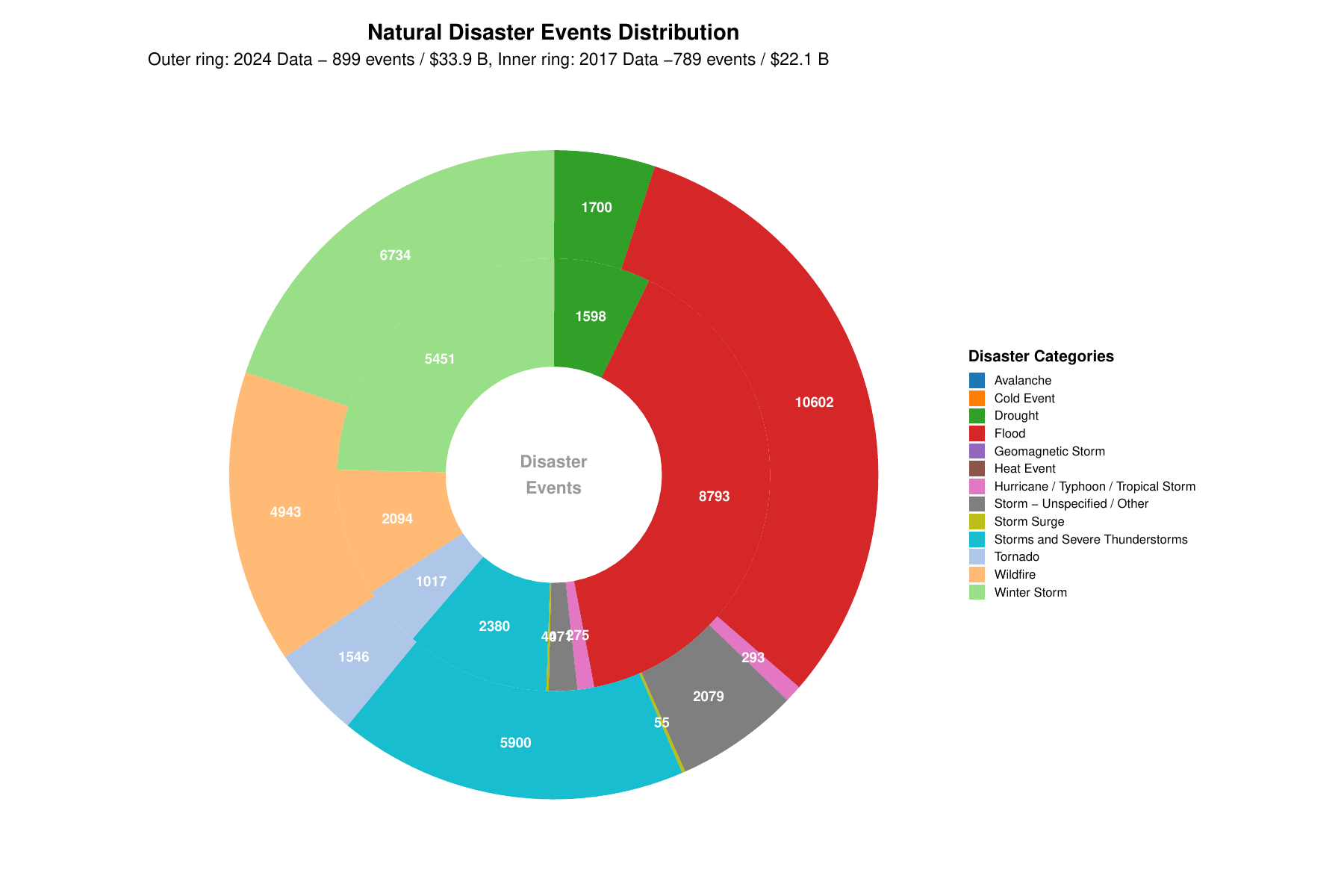}
\end{figure}

Losses between 2024 and 2017 are compared in Figure~\ref{dol}. All financial losses\footnote{In both collections, losses have been normalized to Year 2000 using the Consumer Price Index (CPI).} are expressed in Year 2000 dollars. We observe that losses associated with Thunderstorm, Wildfire, and Storm Other\footnote{Later analysis about frequency and severity didn't yield convincing evidence.} have increased substantially in relative weight. For example, comparing 2024 with 2017:
\begin{itemize}
    \item Thunderstorm: 17.4\% vs. 11\%
    \item Wildfire: 15\% vs. 10\%
\end{itemize}

As mentioned earlier, the frequency of events has remained largely stable between the two collections. The observed increase in loss weights therefore implies that the severity—or loss per event—has risen. We will further explore this relationship in Section~\ref{ms}. This trend also suggests that Canada is becoming both hotter and wetter, as a warming climate seem to contribute to more severe damage from wildfires and a variety of storm types across all seasons except winter.

\begin{center}
\fbox{\begin{minipage}{30em}
\textbf{Top Natural Disasters in Canada: All Time}\\
\textbf{By frequency:} Flood, Thunderstorm, Wildfire, Winter Storm, Tornado, and Drought.\\
\textbf{By total losses:} Flood, Winter Storm, Thunderstorm, Wildfire, Storm Other, and Drought.
\end{minipage}}
\end{center}

\section{Frequency and Severity}\label{fs}

One key goal of this paper is to generate the aggregate annual loss distribution for all event types. We achieve the goal by two steps:
\bi
\item \textbf{Step 1:} Study and simulate annual loss distribution for individual event types. We follows a simple structure: the number of events in a year follows a Poisson distribution, while the individual event loss follows an extreme value distribution (e.g., the Generalized Pareto Distribution, GPD). We are interested in the behaviour of the sum of individual annual losses for the same event, whose overall distribution is a compound function of these two distributions. Step 1 is implemented in Sections~\ref{mf},~\ref{ms}, and~\ref{is}.

\item \textbf{Step 2:} Aggregate annual losses across all event types by Copula. Step 2 is implemented in Section \ref{aal}. 

\ei

\subsection{Modelling of Frequency}\label{mf}

We model the frequency of natural disasters using the Poisson distribution, which assumes that the occurrence of each event is independent of others.

\begin{table}[h!]
\caption{Number and Frequency of Disasters: Non-zero Losses}\label{freq1}
\centering
\begin{tabular}{lrrrrr}\\\hline
Type & \# in 2017 & Freq ($\lambda$): \vline & \# in 2024 & Freq ($\lambda$): \vline & Wald \\
 & Collection & 1955--2016 \vline & Collection & 1955--2020 \vline & Test \textit{t} \\ \hline
Avalanche & 2 & 0.05 \vline & 2 & 0.03 \vline & 0.51 \\
Cold Event & 3 & 0.08 \vline & 3 & 0.05 \vline & 0.79 \\
Drought & 5 & 0.08 \vline & 5 & 0.08 \vline & 0.10 \\
Flood & 150 & 2.46 \vline & 155 & 2.38 \vline & 0.27 \\
Hurricane / Typhoon / &&&&&\\
Tropical Storm & 11 & 0.18 \vline & 11 & 0.17 \vline & 0.15 \\
Storm -- Other & 14 & 0.23 \vline & 21 & 0.32 \vline & -1.00 \\
Storm Surge & 7 & 0.11 \vline & 7 & 0.11 \vline & 0.12 \\
Thunderstorm & 81 & 1.33 \vline & 88 & 1.35 \vline & -0.13 \\
Tornado & 20 & 0.33 \vline & 21 & 0.32 \vline & 0.05 \\
Wildfire & 26 & 0.43 \vline & 27 & 0.42 \vline & 0.09 \\
Winter Storm & 16 & 0.26 \vline & 23 & 0.35 \vline & -0.93 \\ \hline
Grand Total & 335 & 5.49 \vline & 363 & 5.58 \vline & -0.22 \\ \hline
\end{tabular}
\end{table}

Table~\ref{freq1} summarizes the total number of events by disaster types and the estimated frequency for all disaster types in the 2017 and 2024 collections. Only events with non-zero losses are tracked and modeled. Under the Poisson framework, the frequency parameter $\lambda$ is calculated as the total number of events in the sample period divided by the number of years in that period. For example, in the 2017 collection, there were 150 flood events recorded between 1955 and 2016. Therefore, the estimated frequency parameter is $\lambda = 150 / (2016 - 1955) = 2.46$. In both collections, 1955 is the first year in which a non-zero loss event was recorded.

In the 2024 collection, the highest $\lambda$ values are observed for the following categories: Flood, Thunderstorm, Wildfire, Winter Storm, Tornado, and Storm Other (in decreasing order). This ordering, based on the 1955--2020 period, aligns closely with the top six categories based on all recorded events\footnote{Including those with zero losses.} from 1900 to 2020. The only exception is Drought, which has declined in relative importance in recent decades. Conversely, Storm Other have become more frequent, roughly on par with Tornadoes. This provides further evidence that Canada has become wetter over the past 65 years.

When comparing $\lambda$ values between the two collections, they appear quite similar. This finding aligns with the results in Section~\ref{trend}, indicating no strong evidence that any type of natural disaster has become more frequent nationwide in the last seven years. Indeed, the \textit{Wald} test statistics ($t$) are not significant, confirming that the observed differences are statistically negligible.

The estimated frequency parameters ($\lambda$) will be used in the subsequent simulation exercises to generate annual loss distributions for each event type. It is also worth noting that for certain disaster types (e.g., Avalanche, Cold Event, Drought, Hurricane/Typhoon/Tropical Storm, and Storm Surge), the number of events with non-zero losses is limited. For these types, we do not model frequency and severity separately. Instead, we use the simple annual average loss as an estimate when aggregating annual losses across all event types. 

\begin{center}
\fbox{\begin{minipage}{30em}
\textbf{Most Frequent Types of Natural Disasters with Non-zero Losses: 1955--2020}\\
Top six by frequency: Flood, Thunderstorm, Wildfire, Winter Storm, Tornado, and Storm Other. Drought has become less frequent and less impactful during this period, further supporting evidence that Canada is becoming wetter. 
\end{minipage}}\end{center}
\subsection{Modeling of Severity}\label{ms}

The severity of disasters is modeled using the Generalized Pareto Distribution (GPD). As demonstrated by \cite{Gne}, for a wide range of underlying distributions, the conditional distribution of extreme values follows a GPD. The GPD is defined as

\begin{equation}
G_{\xi, \beta}(y) = 1 - (1 + \xi \frac{y}{\beta})^{-1/\xi},
\end{equation}

where $y$ represents the variable taking on extreme values (i.e., exceedances over a threshold, here zero dollar losses), and $\xi$ and $\beta$ are the parameters to be estimated. The shape parameter $\xi$ determines the heaviness of the tail of the distribution, while the scale parameter $\beta$ governs the spread of the data.

The parameters $\xi$ and $\beta$ in the GPD are estimated using all non-zero loss events for each disaster type since 1955. Estimation is performed by maximum likelihood, given that the probability density function is known. All statistical computations are conducted in \textsf{R}. The parameter estimates from the 2017 and 2024 collections are presented in Table~\ref{tab:example}. As noted in Section~\ref{trend}, Thunderstorm, Wildfire have shown increased severity. The results below confirm that the scale parameters for Thunderstorm and Wildfire have increased substantially between the two periods.

\begin{table}[h!]
  \centering
  \caption{Estimation of $\xi$ and $\beta$ in the GPD Based on the 2017 and 2024 Collections}
  \label{tab:example}
  \begin{tabular}{|l|cc|cc|}
    \hline
    Event Type & 2017 $\xi$ & 2017 $\beta$ & 2024 $\xi$ & 2024 $\beta$ \\ \hline
    Flood & 1.180679 & 07.497537 & 1.124818 & 10.276821 \\ \hline
    Winter Storm & 2.304854 & 06.285881 & 1.266917 & 30.233017 \\ \hline
    \textbf{Thunderstorm} & 0.666664 & 13.477403 & 0.635441 & 26.723741 \\ \hline
    \textbf{Wildfire} & 2.568502 & 03.041375 & 2.675231 & 03.893926 \\ \hline
    Storm -- Other & 1.828732 & 05.264203 & 0.282171 & 72.558157 \\ \hline
    Tornado & 1.482868 & 09.622024 & 1.249371 & 16.206468 \\ \hline
  \end{tabular}
\end{table}

\begin{figure}[h!]
\caption{Wildfire Severity Fit Based on GPD}\label{wf}
\begin{subfigure}[b]{0.4\textwidth}
\centering
\includegraphics[width=.7\textwidth]{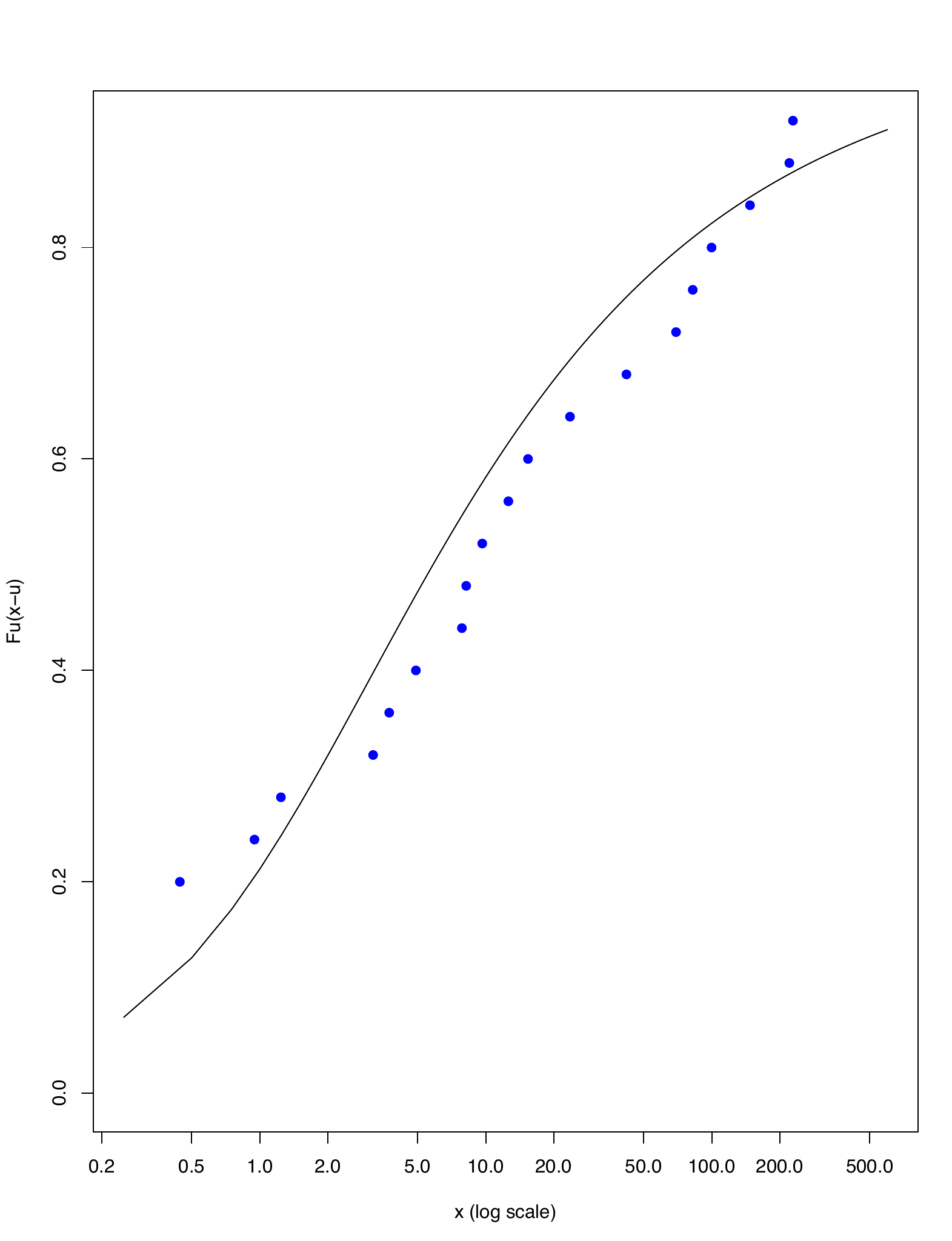}
\caption{2017 collection}
\end{subfigure}
\hfill
\begin{subfigure}[b]{.4\textwidth}
\centering
\includegraphics[width=.7\textwidth]{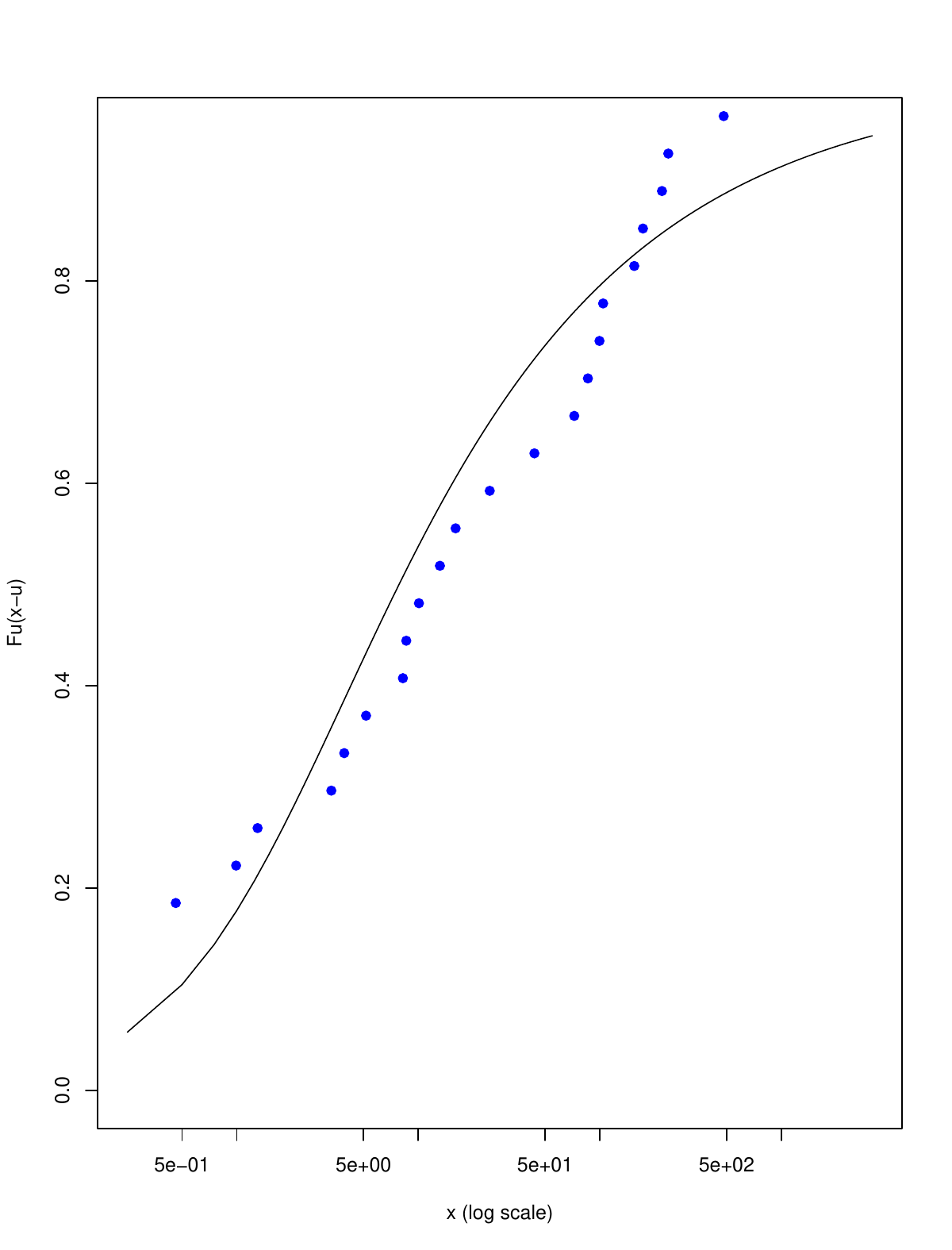}
\caption{2024 collection}
\end{subfigure}
\end{figure}

We now compare empirical fits from the 2017 and 2024 collections to assess whether the severity of Thunderstorm and Wildfire has indeed increased:
\bi
\item \textbf{Wildfire:} Wildfire exhibit a clear increase in severity. As shown in Figure~\ref{wf}, the $x$-axis scale approximately doubled in 2024 compared to 2017, and large-loss events became much more frequent. For example, in 2017, a loss exceeding \$437~MM corresponded to a one-in-10-year event (the 90th percentile). In 2024, that threshold rose to \$687~MM. Table~\ref{tab:example} confirms that both the tail parameter $\xi$ and the scale parameter $\beta$ increased between the two periods. This pattern is expected: the magnitude of fire-related losses has intensified. The largest loss in the 2024 collection was the 2016 Fort McMurray wildfire, approximately \$3.2~B. In comparison, the largest loss in the 2017 collection was the Slave Lake fire in Alberta (2011), around \$600~MM.

\item \textbf{Thunderstorm:} The 2024 collection contains a greater number of high-loss events (blue dots) clustered in the upper-right region, suggesting a modest increase in extreme outcomes. However, the largest loss in both datasets corresponds to the same event: the 1991 thunderstorm in Calgary, Alberta. 

\begin{figure}[h!]
\caption{Thunderstorm Severity Fit Based on GPD}\label{th}
\begin{subfigure}[b]{0.4\textwidth}
\centering
\includegraphics[width=.7\textwidth]{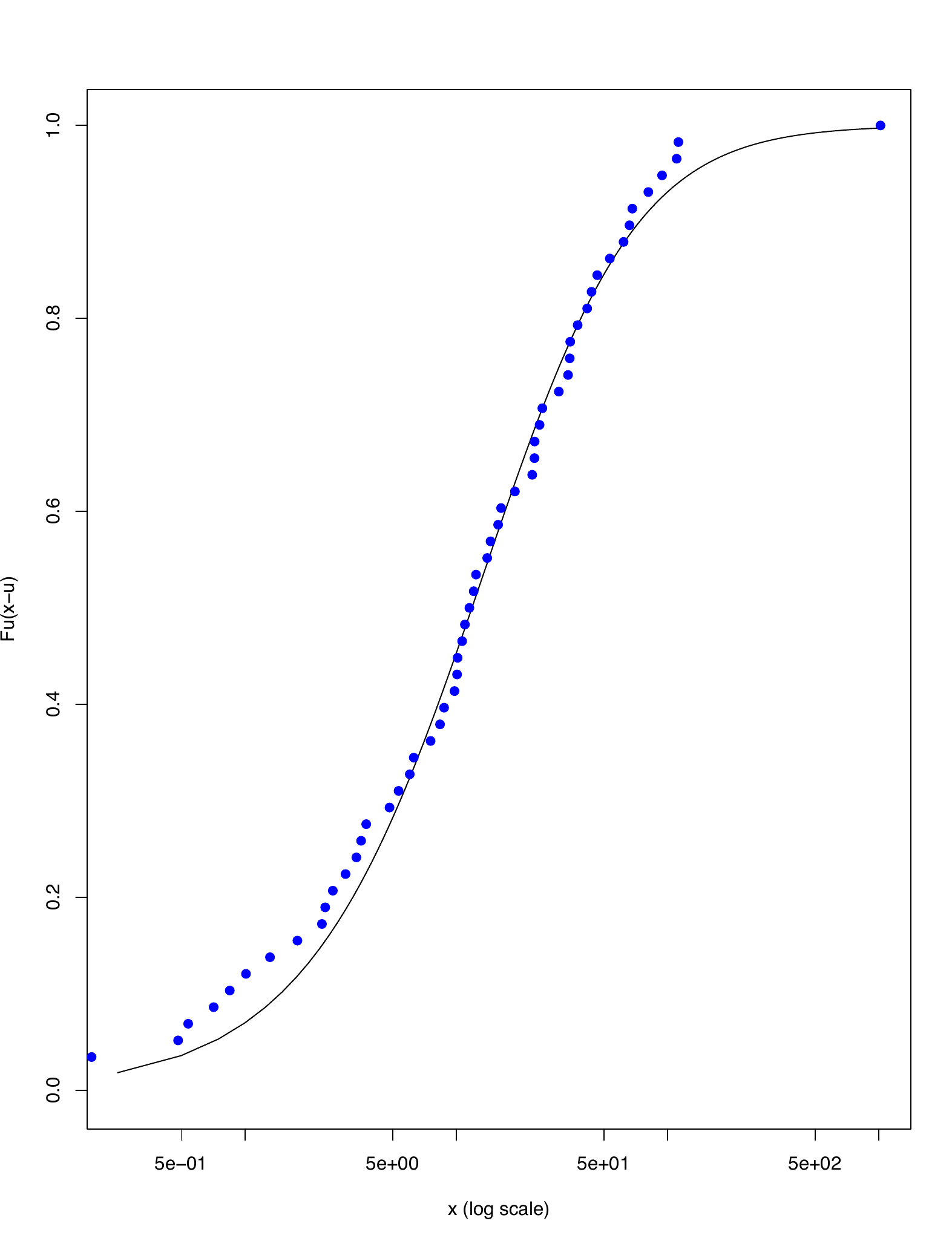}
\caption{2017 collection}
\end{subfigure}
\hfill
\begin{subfigure}[b]{.4\textwidth}
\centering
\includegraphics[width=.7\textwidth]{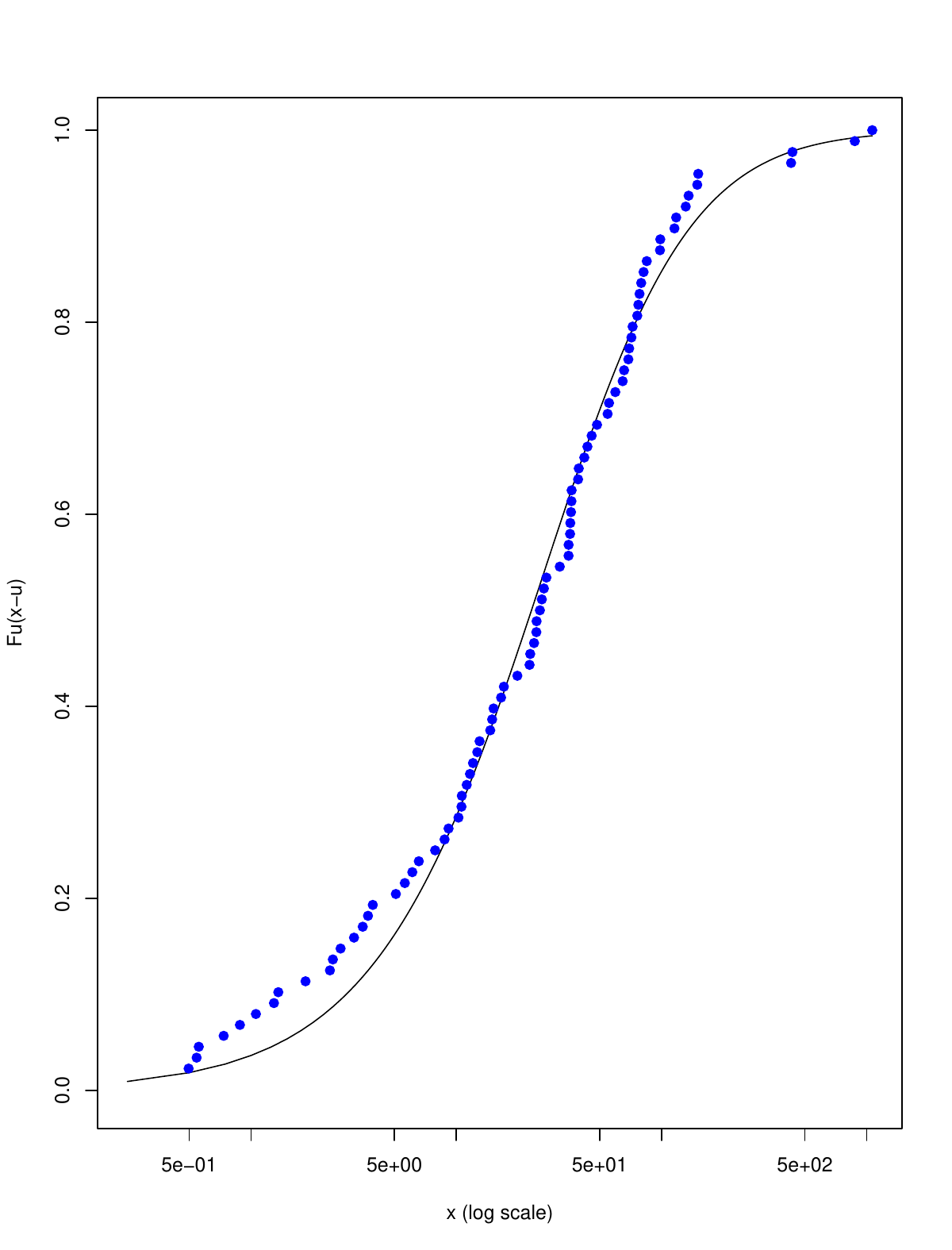}
\caption{2024 collection}
\end{subfigure}
\end{figure}
\ei

\section{Simulation of Annual Losses for Each Disaster Type and Their Aggregate}
\subsection{Simulation of Annual Losses for Each Disaster Type}\label{is}

With both frequency and severity parameters estimated, we simulate the annual loss distribution for each disaster type. In each simulation, for a given disaster type, we first simulate the number of events in a year based on a Poisson distribution, and then the dollar losses of these events using a Generalized Pareto Distribution (GPD). We then sum these simulated losses to obtain the total loss for one year for one disaster type. This process is repeated many times to generate a stable empirical distribution.

Although this simulation approach is intuitive and straightforward, it is computationally intensive. In this paper, we adopt the Fast Fourier Transform (FFT) method. The advantage of FFT is that it can produce the annual loss distribution for each disaster type analytically, given the specified frequency and severity parameters. This analytic approach greatly improves computational efficiency, especially when the number of simulations is large. The mathematical background of FFT is described in Section~\ref{mb}.

\begin{figure}[htbp]
\centering
	\begin{subfigure}{0.3\textwidth}
		\centering
		\includegraphics[width=\textwidth]{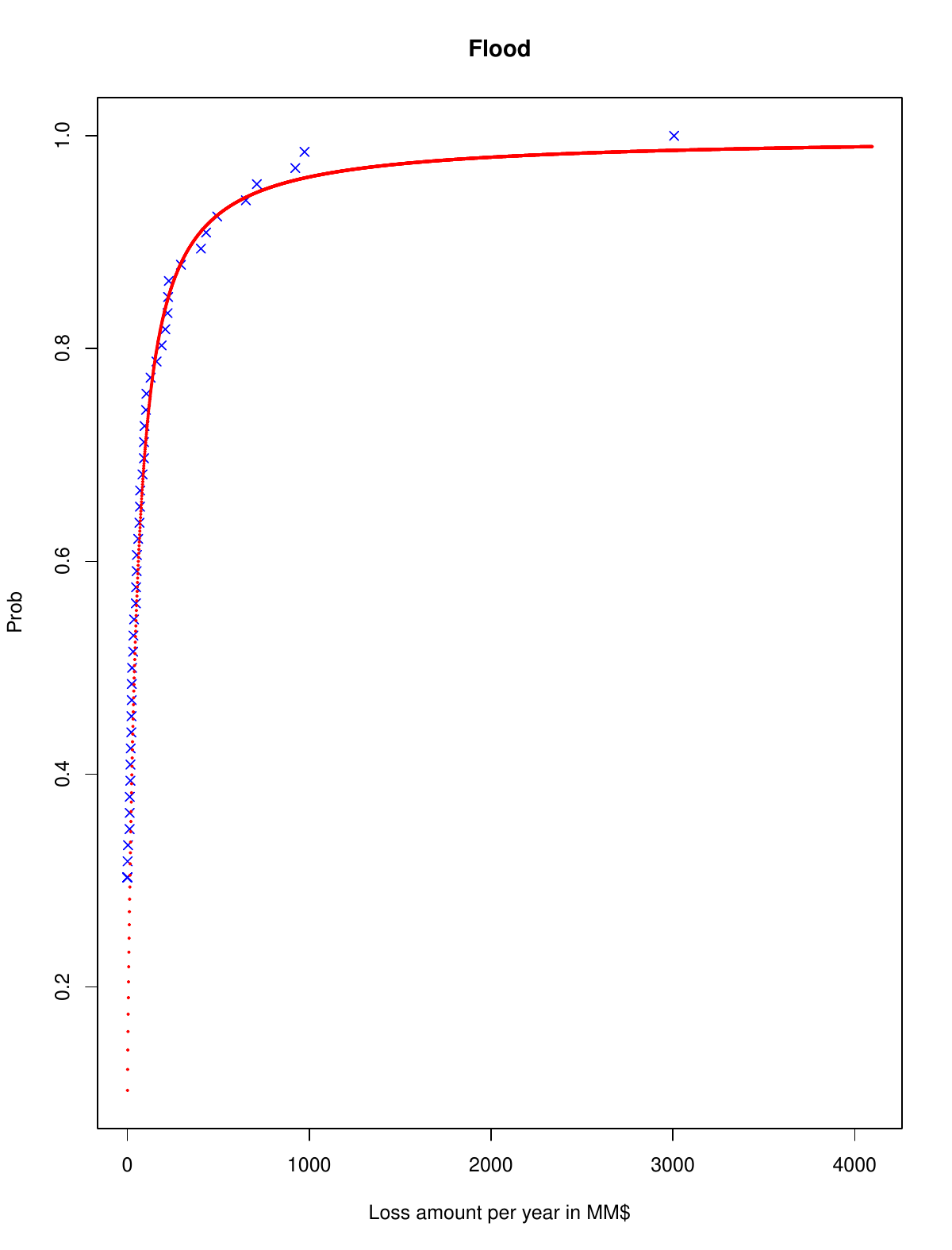}
		\caption{Flood}
		\label{als1}
	\end{subfigure}
	\hfill
	\begin{subfigure}{0.3\textwidth}
		\centering
		\includegraphics[width=\textwidth]{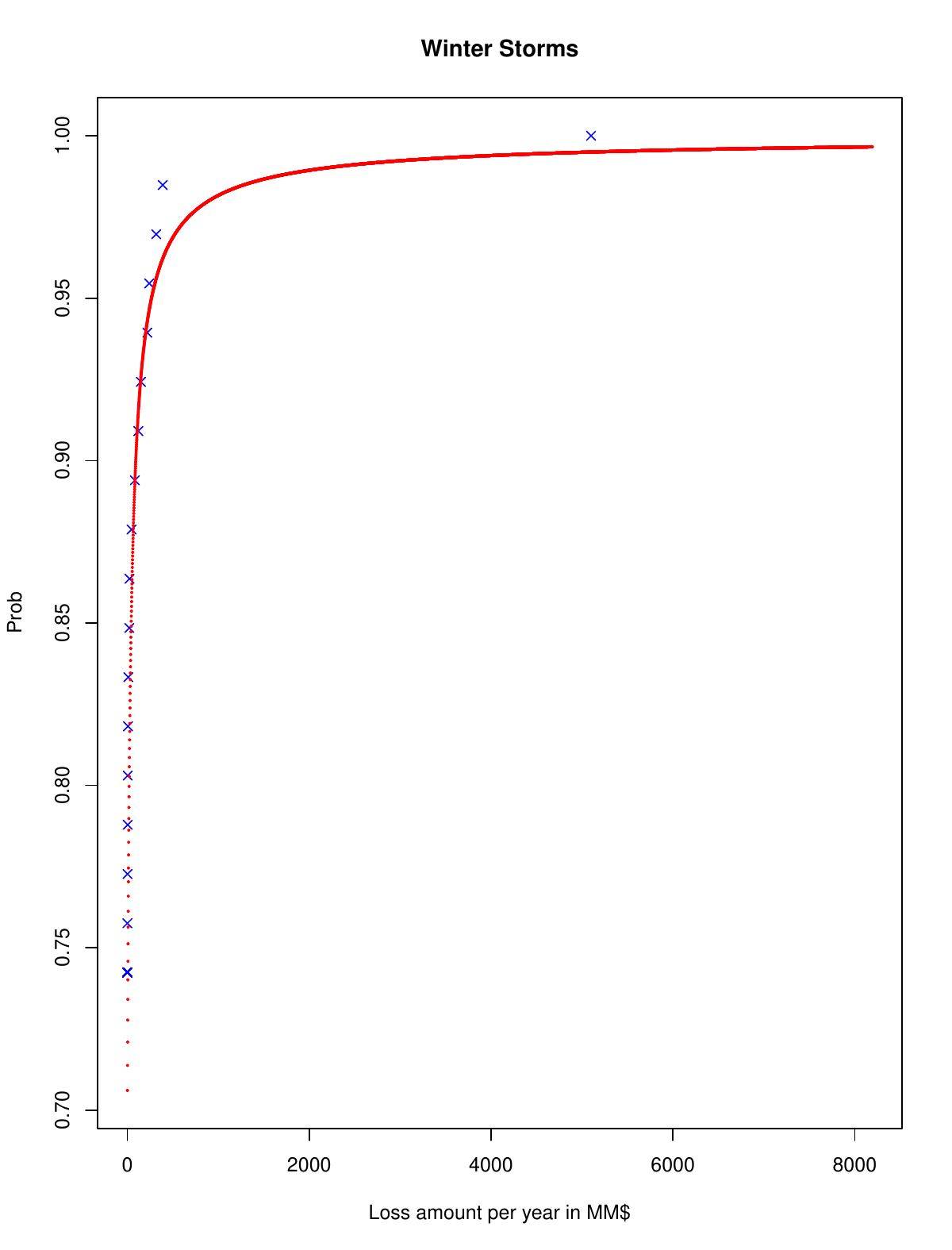}
		\caption{Winter Storm}
	\end{subfigure}
	\hfill
	\begin{subfigure}{.3\textwidth}
		\centering
		\includegraphics[width=\textwidth]{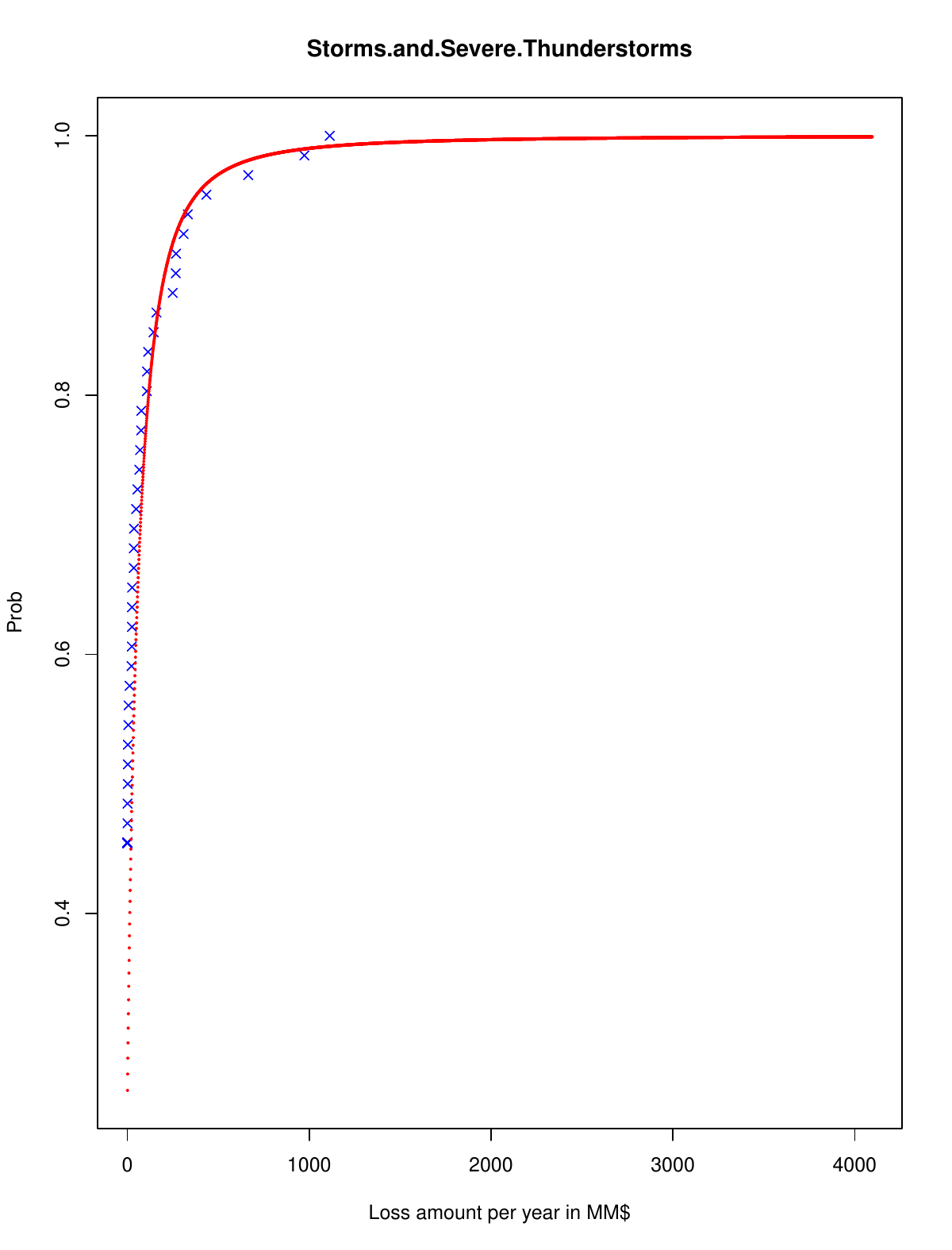}
		\caption{Thunderstorm}
	\end{subfigure}
	\vspace{0.5cm}

	\begin{subfigure}{.3\textwidth}
		\centering
		\includegraphics[width=\textwidth]{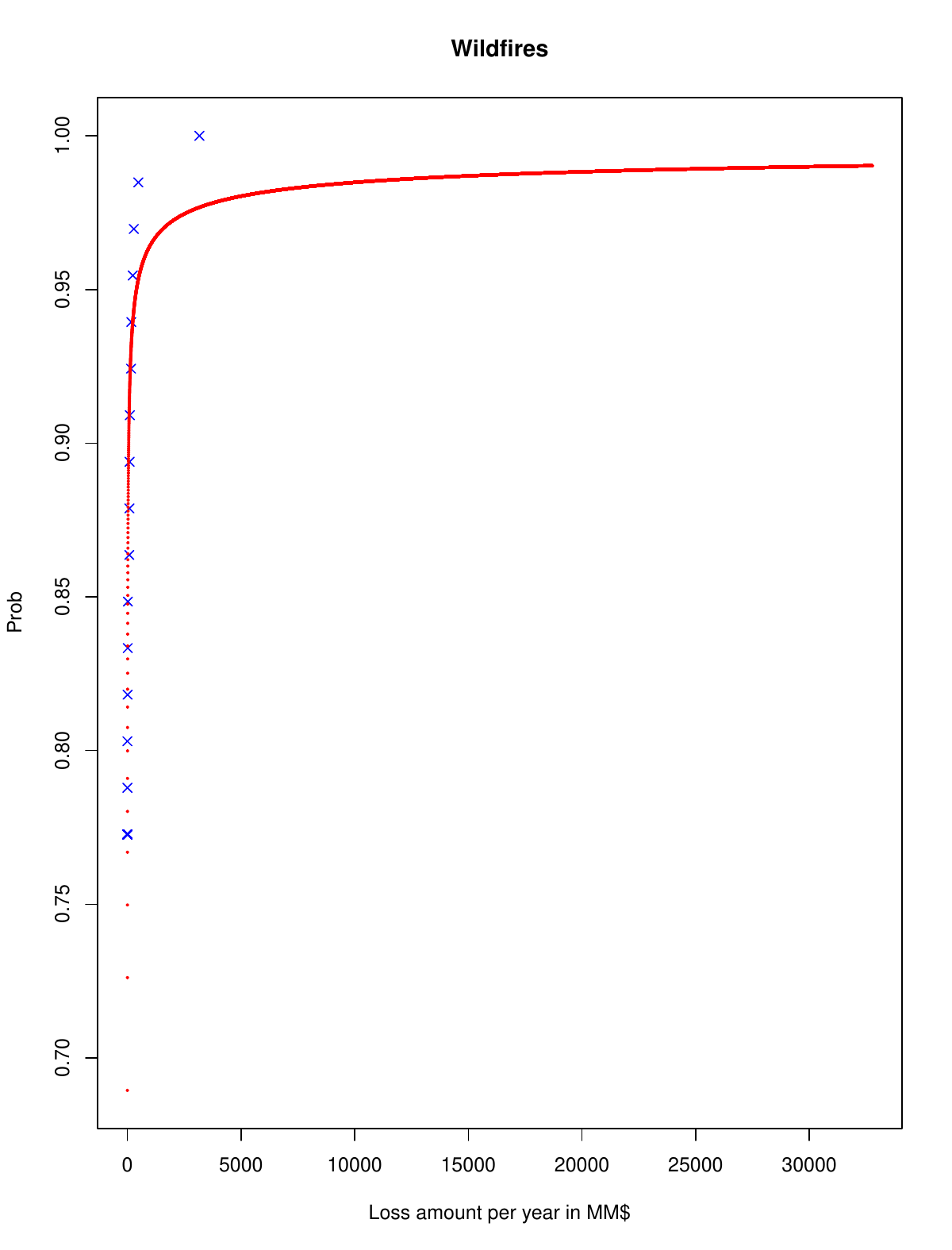}
		\caption{Wildfire}
		\label{als4}
	\end{subfigure}
	\hfill
	\begin{subfigure}[b]{.3\textwidth}
		\centering
		\includegraphics[width=\textwidth]{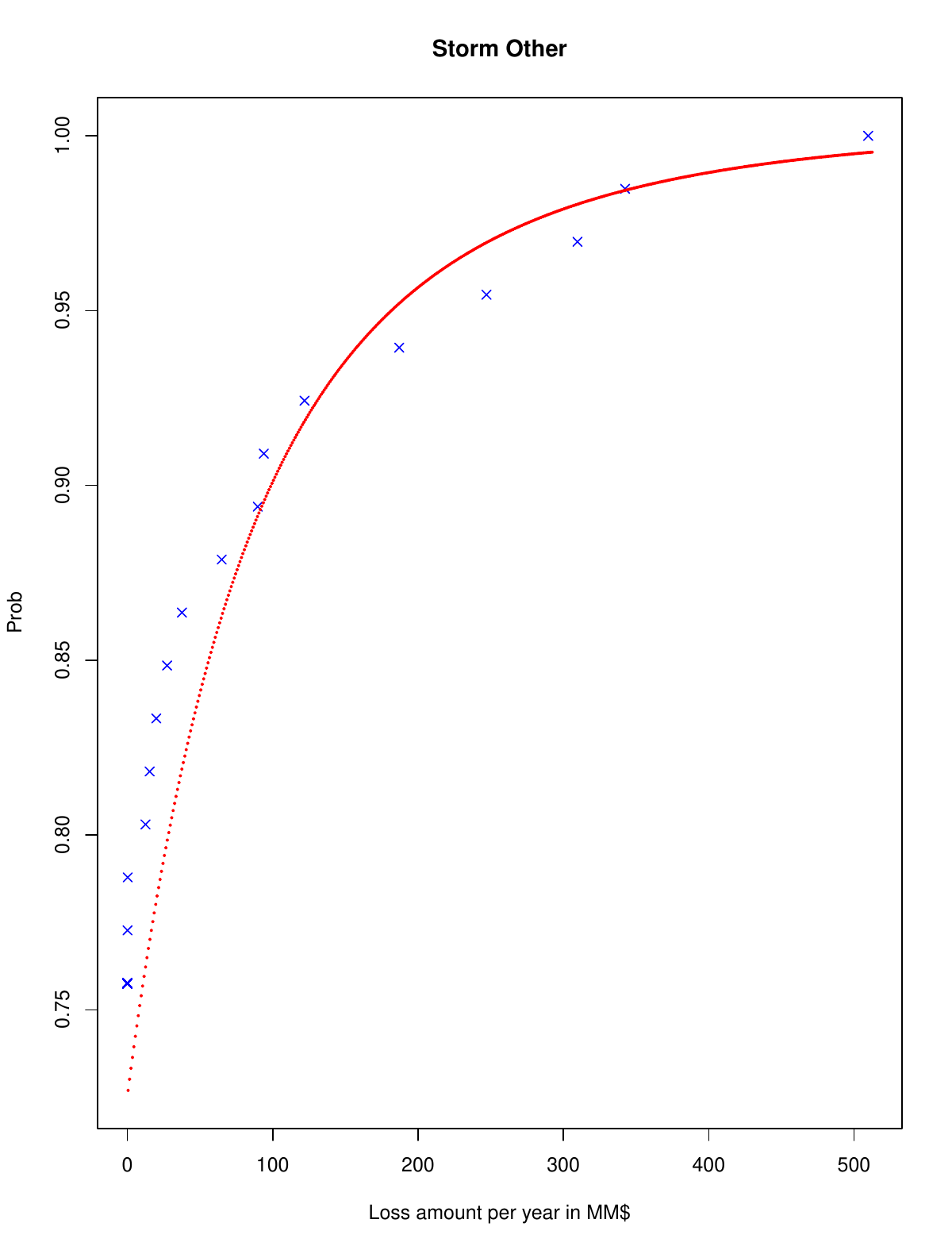}
		\caption{Storm and Other}
	\end{subfigure}
	\hfill
	\begin{subfigure}[b]{.3\textwidth}
		\centering
		\includegraphics[width=\textwidth]{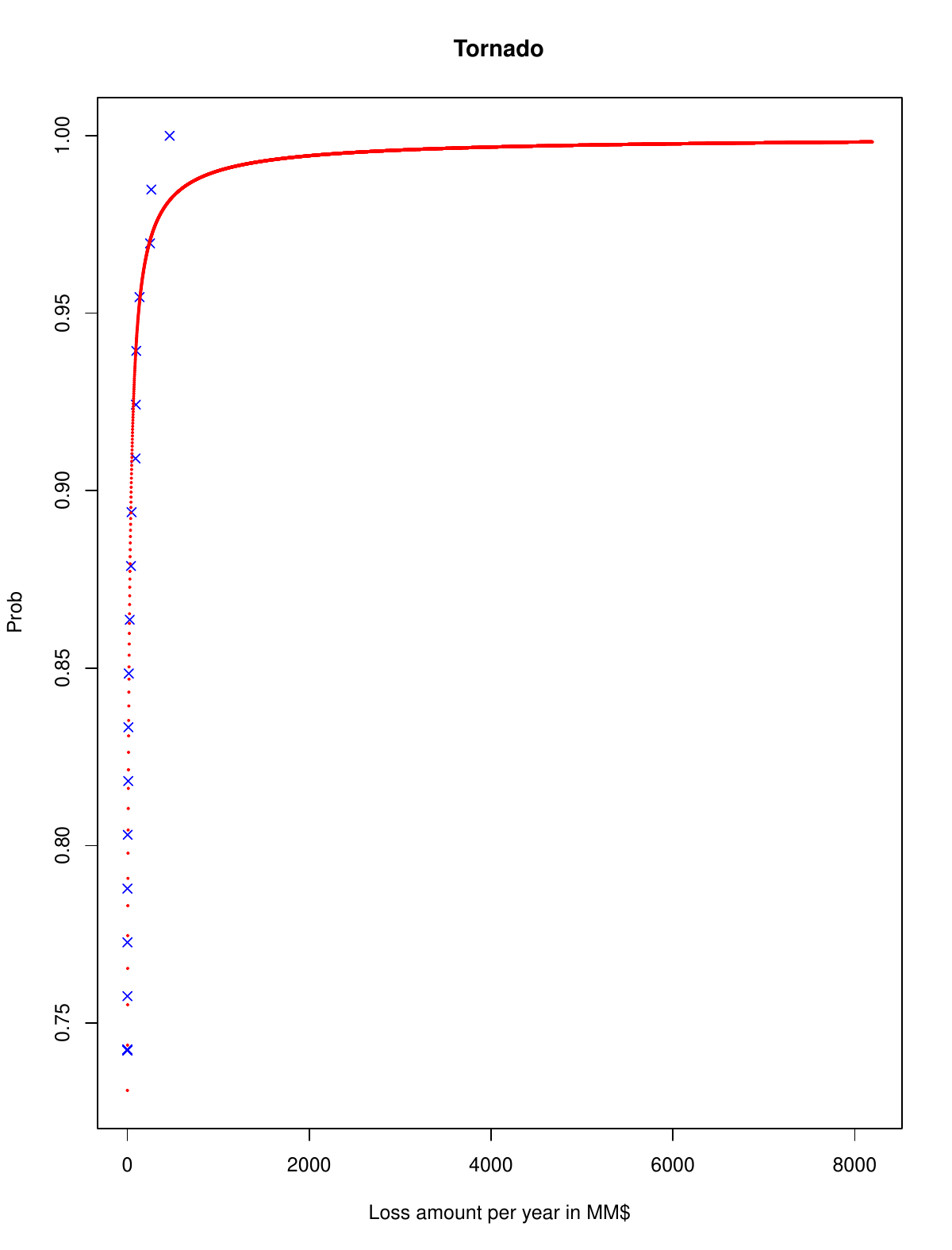}
		\caption{Tornado}
		\label{als6}
	\end{subfigure}
\caption{Annual Loss Simulation by Event Type: 2024 Collection}\label{als}
\end{figure}

Figure~\ref{als} compares the empirical distribution of annual losses for each disaster type (blue) with their FFT-based simulated distributions (red). The main observations are:

\bi
\item \textbf{Which event has the highest actual losses?}\\
Flood and Winter Storm stand out for the magnitude of their annual losses. From Figure~\ref{als1}, the most significant annual flood loss approaches \$3 billion. The high losses result from both frequency (on average more than twice per year, with $\lambda = 2.38$) and the severity of each event. Winter Storm exhibits even higher losses, up to \$5.5 billion per year. Given their much lower frequency (once every three years on average, with $\lambda = 0.35$), this indicates extremely severe individual events.

\item \textbf{Which event has the highest loss potential?}\\
Wildfire (Figure~\ref{als4}) exhibit the longest right tail in their simulated distribution, indicating extreme loss potential. Table~\ref{td} presents the 50th (median), 90th, and 99th percentiles of simulated annual losses by event type. The 99th percentile for Wildfires exceeds \$30.4 billion—about seven times higher than Flood, which have the second-highest 99th percentile at \$4.3 billion.

\item \textbf{Which events have worsened between the 2017 and 2024 collections?}\\
Thunderstorm and Wildfire have become notably more damaging in terms of annual losses. Their tail losses (90th and 99th percentiles) have nearly doubled over this period.
\ei

\begin{table}[h!]
  \centering
  \caption{Annual Loss Simulation by Event Based on 2017 and 2024 Collections}
  \label{td}
  \begin{tabular}{|l|rrr|rrr|}
    \hline
    &&2017&&&2024&\\
    Event Type (\$MM) & 50\% & 90\% & 99\% & 50\% & 90\% & 99\% \\ \hline
    Flood & 32 & 309 & 4,077 & 40 & 362 & 4,331 \\ \hline
    Winter Storm & 0 & 21 & 4,797 & 0 & 93 & 2,157 \\ \hline
    \textbf{Thunderstorm} & 14 & 113 & 543 & 28 & 217 & 981 \\ \hline
    \textbf{Wildfire} & 0 & 46 & 18,346 & 0 & 59 & 30,429 \\ \hline
    Storm and Others & 0 & 10 & 745 & 0 & 100 & 408 \\ \hline
    Tornado & 0 & 31 & 997 & 0 & 43 & 984 \\ \hline
  \end{tabular}
\end{table}

Another observation is that for certain events—such as Winter Storm, Wildfire, Storm and Other, and Tornado—the cumulative distribution function (CDF) of simulated annual losses begins with a relatively high probability at zero losses. This is expected, since these event types occur infrequently (often once every a few years). For example, with $\lambda = 0.3$, the probability of observing zero losses in a given year is approximately 70\%.\footnote{$\lambda$ in Table~\ref{freq1} indicates event frequency.}

\begin{center}
\fbox{\begin{minipage}{30em}
\textbf{Summary:} Loss frequency and severity are modeled separately. Simulated annual losses are generated using FFT.  Thunderstorm and Wildfire are more likely to produce larger extreme losses.
\end{minipage}}\end{center}

\subsection{Aggregation of Annual Losses for All Disaster Types}\label{aal}

As mentioned earlier, our ultimate goal is to estimate the expected total annual loss across different event types. In this section we first generate aggregated simulated losses and compare with empirical annual losses to evaluate model performance. 

When aggregating simulated losss across different event types, one question is how to sum across event types. One conservative approach is to sum all losses from each disaster type, assuming that all losses occur in the same year. This approach likely overestimates aggregate losses because different events might not happen simultaneously. 

Table~\ref{spearman} shows the Spearman correlation among six major disaster types between 1955 and 2020. All coefficients are positive, with the highest correlation at 60\%. This implies that a simple summation of annual losses across event types will tend to overestimate total losses.  

\begin{table}[h!]
\caption{Spearman Correlation: Annual Losses by Event Type}
\centering
\begin{tabular}{lrrrrrr}\label{spearman}\\
\hline
&&Winter&Thunder-&&&\\
 & Flood & Storm & storm & Wildfire & Storm Other & Tornado \\ \hline
Flood & 100\% & 33\% & 60\% & 33\% & 21\% & 38\% \\
Winter Storm & - & 100\% & 30\% & 9\% & 28\% & 7\% \\
Thunderstorm & - & - & 100\% & 20\% & 40\% & 34\% \\
Wildfire & - & - & - & 100\% & 15\% & 18\% \\
Storm Other & - & - & - & - & 100\% & 17\% \\
Tornado & - & - & - & - & - & 100\% \\ \hline
\end{tabular}
\end{table}

We use a Normal copula\footnote{We use the copula approach to implement Step 2 as referred in Section \ref{fs}.} to model correlations among different event types due to its simplicity. Interested readers can refer to \cite{QRM} for a detailed exploration of Normal and other copulas. Using a Normal copula, we simulate marginal cumulative distributions for each event type. Then, for each iteration, we map these marginals to the corresponding annual losses obtained from the FFT simulation in Section~\ref{is}. Summing the losses across all event types produces an aggregated annual loss. This procedure is repeated 10,000 times to generate a stable distribution of total annual losses.

\begin{figure}[h!]
\caption{Simulated and Realized Annual Total Losses Across All Event Types}\label{as}
\begin{subfigure}[b]{0.45\textwidth}
\centering
\includegraphics[width=.7\textwidth]{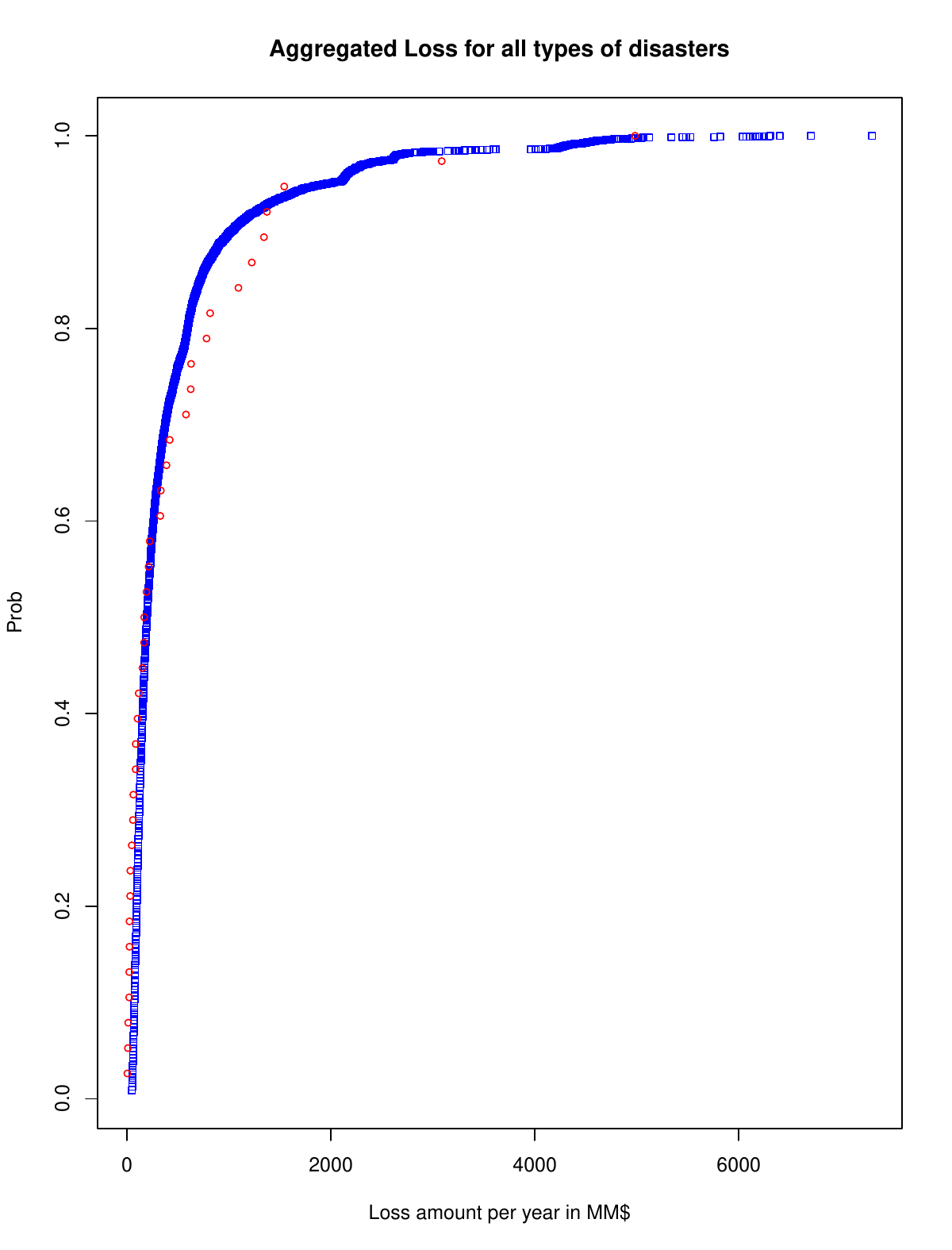}
\caption{2017 Collection}\label{as1}
\end{subfigure}
\hfill
\begin{subfigure}[b]{.45\textwidth}
\centering
\includegraphics[width=.7\textwidth]{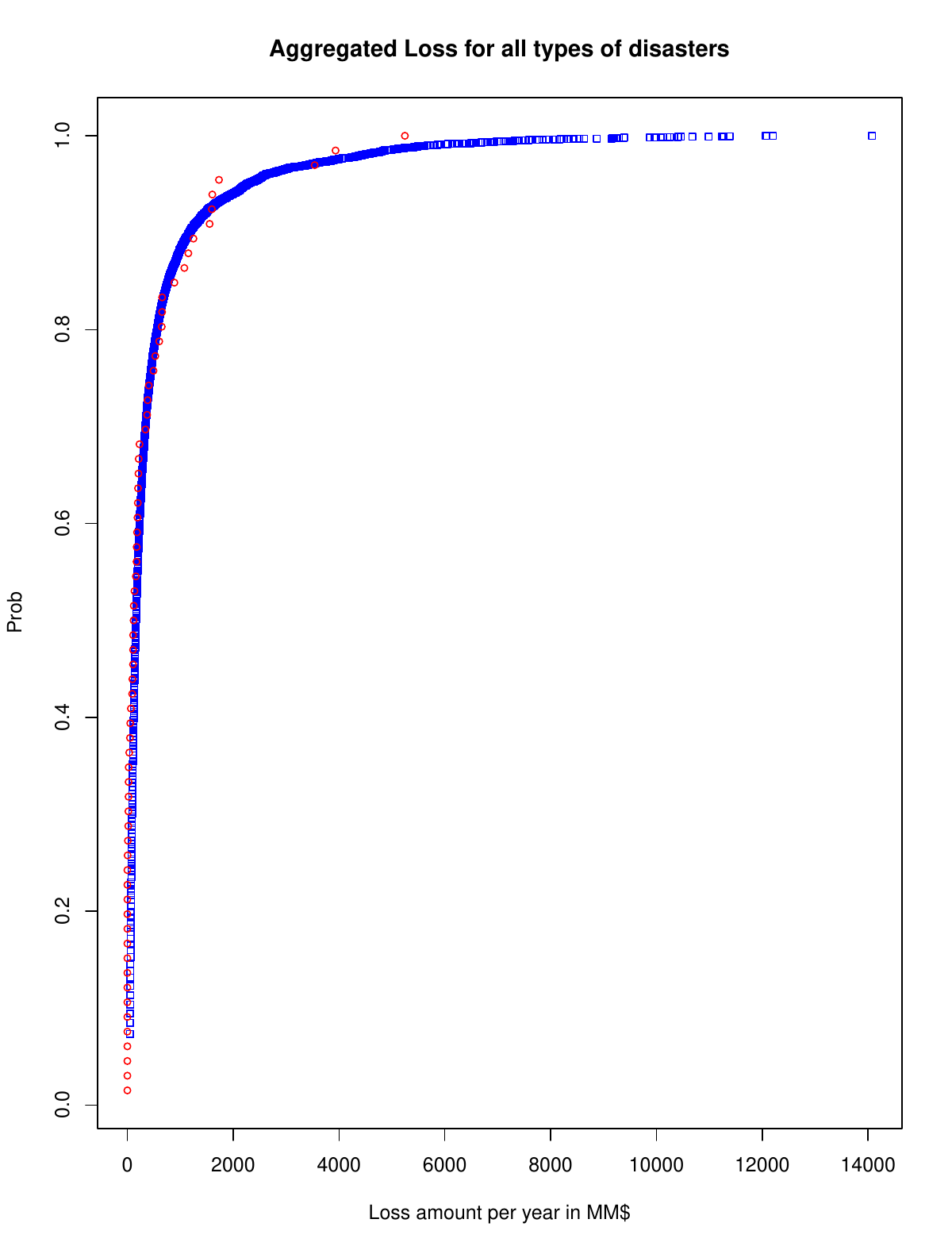}
\caption{2024 Collection}\label{as2}
\end{subfigure}
\end{figure}

In Figure~\ref{as2}, the simulated annual total losses (blue dots) align reasonably well with the actual losses (red dots) for the 2024 collection. Notably, the 2024 data exhibits a longer tail, indicating a higher likelihood of extreme losses. Compared to the 2017 collection (Figure~\ref{as1}), this long-tail effect is more pronounced. Consequently, the main body of the simulated losses for 2024 is lower, which explains why empirical losses move across the simulated losses for higher amounts.

\begin{table}[h!]
\caption{Aggregated Simulated Annual Losses: Medians, Quartiles, and Tail Percentiles}
\centering
\begin{tabular}{rrrrrrr}\label{sumstat}\\\hline
(In \$MM) & 25\% & 50\% & 75\% & 90\% & 99\% & 99.9\% \\ \hline
2017 Collection & 108 & 196 & 478 & 1,012 & 4,312 & 6,038 \\
2024 Collection & \textcolor{red}{78} & \textcolor{red}{164} & \textcolor{red}{432} & \textcolor{green}{1,173} & \textcolor{green}{5,711} & \textcolor{green}{10,460} \\ \hline
\end{tabular}
\end{table}

Table~\ref{sumstat} summarizes the medians, quartiles, and tail percentiles of simulated aggregated annual losses for 2017 and 2024. The 2024 collection shows substantially higher extreme losses (90th, 99th, 99.9th percentiles), while the main body of the distribution (25th, 50th, 75th percentiles) is lower, reflecting a shift toward a heavier tail.

\begin{center}
\fbox{\begin{minipage}{30em}
\textbf{Summary:}  \\
Aggregated annual losses across all disaster types are estimated using a Normal copula. As a country, we are likely to experience larger extreme losses annually.
\end{minipage}}\end{center}

\section{Conclusion}
In this paper, we analyze the Canadian Disaster Database (CDD) to study the frequency and severity of different natural disasters over the past 120 years. Using these data, we generate annual loss distributions for individual disaster types, as well as the total annual loss distribution across all disaster types. Our analysis provides evidence that Canada is experiencing warmer and wetter weather, and suggests a significant likelihood of large extreme losses at the national level.

\renewcommand{\thesection}{Appendix \Roman{section}}
\setcounter{section}{0}
\renewcommand{\theequation}{A-\arabic{equation}}
\setcounter{equation}{0}  
\section{Natural Disasters Database}\label{cdd}

All disaster information was obtained from the Canadian Disaster Database (CDD), maintained by Public Safety Canada. The CDD contains detailed records on over 1,000 natural, technological, and conflict events (excluding war) that have occurred since 1900, either domestically or abroad, and that have directly affected Canadians. 

The database tracks "significant disaster events," defined according to the Emergency Management Framework for Canada, which meet one or more of the following criteria:
\bi
\item 10 or more fatalities
\item 100 or more people affected, injured, infected, evacuated, or rendered homeless
\item an appeal for national or international assistance
\item historical significance
\item significant disruption to normal processes, such that the affected community cannot recover on its own
\ei

CDD records include the location and date of each disaster, the number of injuries, evacuations, and fatalities, as well as approximate financial losses. Data are primarily sourced from reliable and traceable references, including federal institutions, provincial/territorial governments, non-governmental organizations, and media outlets, and are reviewed and updated semi-annually.  

For this study, we focus exclusively on natural disasters related to meteorological or hydrological phenomena. CDD classifies these disasters into the following categories:
\bi
\item Avalanche
\item Cold Event
\item Drought
\item Flood
\item Geomagnetic Storm
\item Heat Event
\item Hurricane / Typhoon / Tropical Storm
\item Storm - Unspecified / Other
\item Storm Surge
\item Storms and Severe Thunderstorms
\item Tornado
\item Wildfire
\item Winter Storm
\ei
\section{Mathematical Background}\label{mb}

\subsection{Overview}
This section summarizes the mathematical and probabilistic framework for modeling the sum of independent individual losses within a year for each event type. We adopt a simple design: the number of events follows a Poisson distribution, and individual losses follow an extreme value distribution (e.g., General Pareto Distribution, GPD). Our interest lies in the behavior of annual losses for each event, which is a compound function of these two distributions. 

Both the Panjer Recursive Method and Fast Fourier Transform (FFT) can be used to study the distribution of total losses (\cite{QRM}, \cite{Panjer}). This paper focuses on FFT, highlighting:
\bi
\item Characteristic function
\item FFT procedure
\item Implementation in R
\ei

\subsection{Model Setup}
Let $K$ denote the (random) number of losses over a fixed time period $[0,t]$, and let $X_1, X_2, \dots$ denote individual losses. The total loss $S$ is
\begin{equation}
S = \sum_{k=1}^{K} X_k
\end{equation}
with $X_i$ independent of $X_j$ for $i \neq j$.  

We assume $K$ follows a Poisson distribution with parameter $\lambda$:
\begin{equation}
f(k) = \frac{e^{-\lambda}\lambda^k}{k!}, \quad k=0,1,2,\dots
\end{equation}
Its probability generating function is
\begin{equation}
P_K(s) = \sum_{k=0}^{\infty} s^k f(k) = \exp[\lambda(s-1)].
\end{equation}

Severity of individual event $X$ is modeled via a Generalized Pareto Distribution (GPD), which captures extreme values according to \cite{Gne}:
\begin{equation}
F^{GPD}_{\xi, \beta}(x) = 1 - \Big(1 + \xi \frac{x}{\beta}\Big)^{-1/\xi},
\end{equation}
where $x$ is the exceedance over threshold (here set to zero), $\xi$ is the shape parameter, and $\beta$ is the scale parameter.

\subsection{Characteristic Function of Total Loss $S$}
The characteristic function of $S$ is
\begin{equation}
\phi_S(z) \equiv \mathbb{E}[e^{i z S}] = P_K(\phi_X(z)), \quad i = \sqrt{-1},
\end{equation}
where $\phi_X(z)$ is the characteristic function of individual losses $X$. Explicitly:
\begin{align}
\mathbb{E}[e^{i z S}] &= \mathbb{E}\Big[e^{i z \sum_{i=1}^{K} X_i}\Big] \nonumber\\
&= \sum_{k=0}^{\infty} \mathbb{E}[e^{i z \sum_{i=1}^k X_i} | K=k] f(k) \nonumber\\
&= \sum_{k=0}^{\infty} (\mathbb{E}[e^{i z X}])^k f(k) \nonumber\\
&= \sum_{k=0}^{\infty} (\phi_X(z))^k f(k) \nonumber\\
&= \exp[\lambda(\phi_X(z) - 1)] \nonumber\\
&= P_K(\phi_X(z)).
\end{align}

\subsection{FFT Procedure}
Following \cite{Panjer}, the FFT maps a probability function to its characteristic function and inverse FFT does vice versa.  
\begin{table}[h!]
\centering
\caption{Summary of FFT Application to $S$}
\begin{tabular}{|c|l|c|l|}
\hline
r.v. & CDF & FFT & Characteristic Function $\phi$ \\ \hline
$X$ & Step 1: $F(X) \rightarrow f(X)$ by discretization & $\Rightarrow$ & Step 2: $\phi_X(z)$ \\ \hline
$S$ & Step 4: $F(S) \leftarrow f(S)$ by cumulation & $\Leftarrow$ & Step 3: $\phi_S(z) = P_K(\phi_X(z)) $ \\ 
&&&$= \exp[\lambda(\phi_X(z)-1)]$\\\hline
\end{tabular}
\end{table}

\bi
\item Step 1: Discretize the severity distribution $F(X)$ to obtain $f_X(0), f_X(1), \dots, f_X(2^r-1)$, with $r$ large enough for accurate approximation.
\item Step 2: Apply FFT to the discretized values to compute $\phi_X(z)$, a vector of $2^r$ complex numbers.
\item Step 3: Compute $\phi_S(z) = P_K(\phi_X(z))$, also a vector of $2^r$ values.
\item Step 4: Apply inverse FFT to $\phi_S(z)$, divide by $2^r$, and cumulatively sum to obtain $F(S)$.
\ei

\subsection{Implementation in \texttt{R}}
Example R implementation for GPD and FFT:
\begin{lstlisting}[language=R]
xi <- 1.180679
beta <- 7.497537
loc <- 0
lambda <- 3.2821
nsteps <- 2^12

svrsmpl <- function(svr, location, xi_input, beta_input) {
  out_value <- numeric(length(svr))
  for (i in 1:length(svr)) {
    quant <- function(pp, xi, beta, u) {
      (1 - (1 + xi*(pp-u)/beta)^(-1/xi))
    }
    out_value[i] <- quant(svr[i], xi_input, beta_input, location)
  }
  return(out_value)
}

# Step 1: Discretization
pts <- c(0, 1:(nsteps+1)-0.5)
f <- diff(svrsmpl(pts, loc, xi, beta))

# Step 2: Characteristic function of X
f.hat <- fft(f)

# Step 3: Characteristic function of S
gf.hat <- exp(lambda*(f.hat - 1))

# Step 4: Obtain CDF of S
s <- Re(fft(gf.hat, inverse=TRUE)/nsteps)
S.df <- cumsum(s)
\end{lstlisting}


\begin{thebibliography}{99}
\bibitem{Gne}
B. Gnedenko, \textit{Sur la distribution limite du terme maximum d'une série aléatoire}, Annal of Mathematics, 1943
\bibitem{QRM} 
A.J. McNeil, R. Frey, P. Embrechts 
\textit{Quantitative Risk Management}. 
Princeton University Press, 2005
\bibitem{Panjer} 
Harry H. Panjer, \textit{Operational Risk: Modeling Analytics}, Wiley INTERSCIECE, 2006 

\end{thebibliography}
\end{document}